\input harvmac
\noblackbox

\lref\hanany{A.~Hanany and E.~Witten,
 Nucl.\ Phys.\ {\bf B492}, 152 (1997) [arXiv:hep-th/9611230].}

\lref\gutowski{J.~Gutowski,
 Nucl.\ Phys.\ {\bf B627}, 381 (2002) [arXiv:hep-th/0109126].}

\lref\cvetic{M.~Cvetic and H.~Soleng,
 Phys.\ Rept.\ {\bf 282}, 159 (1997) [arXiv:hep-th/9604090].}

\lref\bdthm{T. Banks and L. Dixon,
 Nucl.\ Phys.\ {\bf B307}, 93 (1988).}

\lref\contrans{B.~Greene, D.~Morrison and A.~Strominger,
 Nucl.\ Phys.\ {\bf B451}, 109 (1995) [arXiv:hep-th/9504145]\semi
A.~Avram, P.~Candelas, D.~Jancic and M.~Mandelberg,
 Nucl.\ Phys.\ {\bf B465}, 458 (1996) [arXiv:hep-th/9511230]\semi
T.~Chiang, B.~Greene and M.~Gross,
 Nucl.\ Phys.\ Proc.\ Suppl.\ {\bf 46}, 82 (1996) [arXiv:hep-th/9511204].}

\lref\porferr{S.~Ferrara and M.~Porrati,
Phys.\ Lett.\ {\bf B423}, 255 (1998) [arXiv:hep-th/9711116].}

\lref\cchange{S.~Kachru and E.~Silverstein,  
 Nucl.\ Phys.\ {\bf B504}, 272 (1997) [arXiv:hep-th/9704185]\semi
B.~Ovrut, T.~Pantev and J.~Park,
 JHEP {\bf 0005}, 045 (2000) [arXiv:hep-th/0001133].}

\lref\kst{S.~Kachru, M.~Schulz and S.~Trivedi,  [arXiv:hep-th/0201028].}

\lref\drs{K.~Dasgupta, G.~Rajesh and S.~Sethi, JHEP {\bf 9908}, 023 (1999) 
[arXiv:hep-th/9908088].}
 
\lref\seven{J.~de Boer, R.~Dijkgraaf, K.~Hori, A.~Keurentjes, 
J.~Morgan, D.~R.~Morrison and S.~Sethi, 
Adv.\ Theor.\ Math.\ Phys.\ {\bf 4}, 995 (2002) [arXiv:hep-th/0103170].}

\lref\banks{T.~Banks,
[arXiv:hep-th/0011255].}

\lref\GVW{S.~Gukov, C.~Vafa and E.~Witten,
 Nucl.\ Phys.\ {\bf B584}, 69 (2000) [arXiv:hep-th/9906070].}

\lref\SVW{S.~Sethi, C.~Vafa and E.~Witten,
 Nucl.\ Phys.\ {\bf B480}, 213 (1996) [arXiv:hep-th/9606122].}

\lref\KPV{S.~Kachru, J.~Pearson and H.~Verlinde,
 [arXiv:hep-th/0112197].}

\lref\JoeTwo{J.~Polchinski,
 ``String Theory. Vol.~2: Superstring Theory And Beyond,''
 {\it  Cambridge, UK: Univ.\ Pr.\ (1998) 531 p}.}

\lref\joefrey{A.~Frey and J.~Polchinski, [arXiv:hep-th/0201029].}

\lref\GKP{S.~B.~Giddings, S.~Kachru and J.~Polchinski,
[arXiv:hep-th/0105097].}

\lref\GPOne{M.~Grana and J.~Polchinski,
[arXiv:hep-th/0106014].}

\lref\BB{K.~Becker and M.~Becker, 
Nucl.\ Phys.\ {\bf B477}, 155 (1997) [arXiv:hep-th/9605053]}

\lref\Astro{A.~Strominger, 
Phys.\ Lett.\ {\bf B383}, 44 (1996) [arXiv:hep-th/9602111].}

\Title{\vbox{\baselineskip12pt
\hbox{hep-th/0205108}
\hbox{SU-ITP-02/16}
\hbox{SLAC-PUB-9205}
\hbox{TIFR-TH/02-17}
}}
{\vbox{\centerline{Supersymmetry Changing Bubbles}
\vskip2pt\centerline{in String Theory}}}

\centerline{Shamit Kachru $^{a}$\footnote{$^1$}{skachru@stanford.edu},
Xiao Liu $^{a}$\footnote{$^2$}{liuxiao@stanford.edu},
Michael Schulz $^{a}$\footnote{$^3$}{mschulz@stanford.edu} and
Sandip P. Trivedi $^{b}$\footnote{$^4$}{sandip@tifr.res.in}}
\medskip\centerline{$^a$\it Department of Physics and SLAC,
Stanford University}
\centerline{\it Stanford, CA 94305/94309, USA}
\medskip
\centerline{$^b$\it Tata Institute of Fundamental Research}
\centerline{\it Homi Bhabha Road, Mumbai 400 005, INDIA}

\vskip .15in
We give examples of string compactifications to 4d Minkowski space
with {\it different\/} amounts of supersymmetry that can be connected
by spherical domain walls.  The tension of these domain walls is
tunably lower than the 4d Planck scale. The ``stringy'' description of
these walls is known in terms of certain configurations of wrapped
Dirichlet and NS branes.  This construction allows us to connect a
variety of vacua with 4d ${\cal N}=4,3,2,1$ supersymmetry.

\Date{May 2002}

\newsec{Introduction}
String theory is known to have many different vacua.  An important
direction of research aims at understanding whether these different
vacua are connected.  For compactifications to 4 dimensional Minkowski
space the situation is as follows.  With ${\cal N} =4$ supersymmetry
(susy), it is known that there are several disconnected components of
the space of vacua (see e.g.~\seven).  With ${\cal N}=2$ susy, naively
disconnected components are known to be connected up in a large web
\contrans, although it is premature to say that all such models are
connected.  For ${\cal N}=1$, less is known,\foot{In this case, we
generically expect the moduli to be lifted by quantum corrections.
For this reason our discussion will mostly focus on ${\cal N}\geq 2$
vacua, but our idea would also apply to any ${\cal N}=1$ models where
the flux-generated no-scale potential is the full potential.}
although some classical obstructions to connecting vacua are
circumvented by string theory via chirality changing phase 
transitions~\cchange.

In the discussion above, the notion of connectedness relates to moving
along a moduli space of degenerate solutions. It is known that
Minkowski vacua with {\it different} amounts of susy can never be
connected, in this sense.  A theorem to this effect was proved for the
perturbative heterotic theory in \bdthm.  However, it is clear that
weaker notions of connectedness exist and could be physically
relevant.  For instance, two vacua can be connected by a finite
potential barrier, $V_{\rm bar}$.  For $V_{\rm bar}$ much less than
the four dimensional (4d) Planck scale $M_{4}$, low-energy field
theory would correctly describe the dynamics in rolling betwen these
vacua. Such a notion of connectedness might be relevant in cosmology.
A related weaker notion of connectedness requires the existence of
vacuum bubbles of one vacuum inside the other, with the domain wall
separating the two having a tension $\sigma \ll (M_{4})^3$.

In this letter we show that one {\it can} unify some vacua with
different amounts of supersymmetry in this weaker sense.  Our starting
point is IIB string theory compactified on the $T^6/Z_2$
orientifold. This vacuum has ${\cal N}=4$ susy and is a dual
description of the heterotic theory on $T^6$. Appropriately turning on
RR and NS fluxes yields vacua with ${\cal N}= 3,2,1$ susy
\refs{\drs, \kst,\joefrey}.  We show that the vacua with reduced susy can be
connected to the ${\cal N}=4$ vacuum, and to each other, by spherical
domain walls.  In the ten dimensional string theory, these domain
walls are made up of NS and Dirichlet five branes, each of which wrap
an internal three cycle, besides spanning the spherical boundary.  It
is important to note that the tension of the resulting domain walls
can be made parametrically lighter than $(M_{4})^3$ (by tuning the
compactification volume $V$, as we will demonstrate).  This ensures
that the vacuum inside the bubble is not shielded from the one outside
by a black hole horizon, and is available for inspection from the
outside.

The bubble configuration we construct is not BPS and evolves in time,
with a trajectory determined primarily by the tension of the domain
wall.  By tuning the radius of the internal space, one can make the
lifetime of the bubble arbitrarily large.

We should emphasise that the vacua under consideration here are
quantum mechanically stable, and all of them have zero ground state
energy.  As a result, the spherical bubbles referred to above are not
produced by quantum tunneling, as in the decay of a false vacuum.

In \S2, we briefly review the construction of the various vacua in IIB
on $T^6/Z_2$ with fluxes.  \S3 describes the domain wall brane
configurations which interpolate between the different vacua.  We
discuss the construction of the domain walls from wrapped five branes,
their resulting dynamics, and the stability of the walls. As a
concrete example we consider a bubble of the standard ${\cal N}=4$
vacuum inside a theory with ${\cal N}=2$ susy.  We close with a
discussion in \S4.

\newsec{Vacua with various ${\cal N}$ in IIB on $T^6/Z_2$}

Our starting point is IIB theory compactified on a $T^6/Z_2$
orientifold.  This model is T-dual to Type I theory and preserves
${\cal N}=4$ supersymmetry.  $16$ D3 branes are needed to cancel the
RR tadpoles arising from the O3 planes.  The resulting low energy
theory is $SO(32)$ ${\cal N}=4$ supersymmetric Yang-Mills theory
coupled to ${\cal N}=4$ supergravity.

However, this is not the most general possibility.  The IIB
compactification also admits other superselection sectors in which we
turn on quantized fluxes of the three-form field strengths $H$ and $F$
originating from the NS-NS and RR sectors.  That is, $H$ and $F$
satisfy the conditions
\eqn\quant{{1\over {(2\pi)^2 \alpha^\prime}}\int_\gamma F = m_{\gamma}
\in {\bf Z},~~{1\over {(2\pi)^2 \alpha^\prime}} \int_\gamma H = n_{\gamma}
\in {\bf Z}, }
where $\gamma$ labels the classes in $H_{3}(T^6,{\bf Z})$.  For the
case of a six-torus with coordinates $x^i$ and $y^i$, each of period
1, we can be very explicit about this choice.  Let $d\xi^a=dx^i, dy^j,
1\le i,j \le 3$, denote six one-forms. Then, a basis for $H^{3}(T^6,
{\bf Z})$ is given by the twenty three forms, $d\xi^a\wedge
d\xi^b\wedge d\xi^c, 1 \le a,b,c \le 6 $.  For the most general choice
of flux, ${1\over (2 \pi)^2 \alpha^{\prime}}F$ and, ${1\over (2 \pi)^2
\alpha^{\prime}}H$ can be expanded in this basis with integer
coefficients.

In the presence of such fluxes, the full tadpole cancellation
condition for the D3 brane charge reads:\foot{Here we ignore the
possibility of exotic O3 planes and choose the integer coefficients
which characterise the flux to be even, as explained in \joefrey.}
\eqn\tadpole{ {1\over  2 (2\pi)^4 (\alpha^{\prime})^2}\int_{T^6}
H \wedge F~ + N_{D3} ~=~16. } 
Here we consider only the susy preserving case with no anti-branes.
Susy breaking by adding anti-branes and vacuum bubbles in similar
backgrounds was studied in \KPV.

In sectors with non-vanishing flux, one finds an effective
(super)potential for the Calabi-Yau complex structure and K\"ahler
moduli \GVW\ (for a detailed derivation, see Appendix~A of \GKP).
Supersymmetric vacua are located at points in complex structure moduli
space where $G = F - \phi H$ is of type (2,1) (here $\phi$ is the IIB
axio-dilaton), while the K\"ahler structure $J$ should be chosen to
make $G$ primitive (i.e.\ satisfy $J \wedge G =0$).  These conditions
were studied in detail for the case of $T^6/Z_2$ in
\kst, and it was found that for generic choices of the fluxes
there are no supersymmetric critical points.  However, for suitable
non-generic choices of flux, one can find vacua with ${\cal N}=1,2,3$
supersymmetry.  In these vacua, typically all the complex structure
moduli and some of the K\"ahler moduli are fixed.  The dilaton-axion
is also typically fixed with $g_s \sim O(1)$.  One K\"ahler modulus,
governing the overall volume of compactification $V$, is never lifted
in these models; this will be important in the discussion below.

In \S3.3, a specific ${\cal N}=2$ vacuum will be considered. It
corresponds to the choice of flux
\eqn\fluxcho{\eqalign{{1\over (2 \pi)^2 \alpha^{\prime}} F & =
2 dx^1\wedge dx^2\wedge dy^3 + 2 dy^1\wedge dy^2\wedge dy^3 \cr
{1\over (2 \pi)^2 \alpha^{\prime}} H &=
2 dx^1 \wedge dx^2\wedge dx^3 + 2 dy^1 \wedge dy^2 \wedge dx^3.}}

Following \refs{\drs, \kst} one easily finds that there is a moduli space of
${\cal N}=2$ supersymmetric vacua with these fluxes.  A particular
locus in this moduli space has $\phi=i$ and a $T^6$ which is of the
form $(T^2)^3$, where each two-torus has complex structure $\tau = i$.
The K\"ahler form can be chosen to be $J \sim i R^2 \sum_{i=1}^{3}
dz_i \wedge d\bar z_i$.  This is just a product of square two-tori
with overall volume $R^6$.

A quick way to see that the vacuum preserves ${\cal N}=2$
supersymmetry is by noticing that along this locus, $G$ takes the form
\eqn\valG{{1 \over (2 \pi)^2 \alpha^{\prime}} G=-{i \over 2}( dz_1 \wedge d{\bar z}_2 \wedge dz_3 
+ d{\bar z}_1 \wedge dz_2 \wedge dz_3).}  
${\cal N}=2$ susy requires that there be another inequivalent choice
of complex structure which keeps $G$ of type $(2,1)$; this corresponds
to taking $z_{1,2} \rightarrow {\bar z}_{1,2}, z_3 \rightarrow z_3$.

\newsec{Vacuum bubbles from D5 and NS5 branes}

\subsec{Overview}

The key idea in our construction of bubbles is the following: by
wrapping D5/NS5 branes on three cycles of the compact manifold it is
possible to construct domain walls in $R^{3,1}$ across which the
quantised fluxes in the compact manifold jump. E.g., wrapping a D5
brane on a three cycle causes the flux of $F$ through the {\it dual}
three cycle to jump by one unit.  Since the vacua reviewed above
differ essentially in the RR and NS fluxes along the compact
directions, this allows different vacua to be connected.

In fact this idea was used in \GVW\ to construct BPS domain walls
between ${\cal N}=1$ vacua in the setting of non-compact Calabi-Yau
constructions.  Our interest is in compact internal manifolds,
resulting in flat 4d spacetime.  In this case, we do not expect BPS
domain walls to interpolate between vacua with different amounts of
supersymmetry for two reasons.  First, the central extensions of the
supersymmetry algebra do not admit BPS domain walls of nonzero tension
between supersymmetric Minkowski vacua in supergravity (see e.g.\
\refs{\cvetic,\porferr}).\foot{ This is not true in global
supersymmetry.  The additional constraint in supergravity arises
roughly because one needs the superpotential $W$ to vanish for a
Minkowski vacuum.}  Second, planar domain walls have codimension one
and are often singular in supergravity, see e.g.~\gutowski.

With this in mind, we construct non-BPS spherical domain walls in
$R^{3,1}$, separating a bubble of one vacuum inside the wall from
another vacuum outside.  Two requirements must be met by the domain
wall to consistently interpolate between the vacua. First, the flux of
$F,H$ must jump appropriately across the wall.  Second, the moduli
must vary smoothly across it.  It is clear that any jump in $F,H$
fluxes can be engineered by choosing D5,NS5 branes wrapping three
cycles in the appropriate homology classes.  We will choose the
minimum area three cycle in each homology class which is consistent
with our boundary conditions.  The domain wall is then the composite
configuration made out of the resulting D5,NS5 branes.

To meet the condition on the moduli, we restrict ourselves here to
considering pairs of vacua such that moduli lifted in both vacua are
fixed to the same values.  The remaining moduli, unfixed in one or
both vacua, can then simply be tuned to take the same values on both
sides of the wall (we will show that the backreaction of the walls is
small enough to make this a good approximation).

In fact, this condition is not very restrictive, and allows our
construction to connect several vacua, including many with different
susy's.  For example, since none of the moduli in the standard ${\cal
N}=4$ vacuum are fixed, it can be connected to all the other vacua in
the above manner.  This is enough to establish that all the vacua of
\S2 are connected by the above construction.

The two vacua connected by the wall will in general have different
numbers of D3 branes \tadpole. One can verify that the extra D3 branes
in one vacuum terminate on the $5$ branes making up the domain wall
consistently \Astro.  Also, we note that being composed of $5$ branes,
the resulting domain walls have a thickness of order the string
scale. As a result, in analysing their dynamics below we can work in
the thin wall approximation.

Some features of the resulting domain wall dynamics were discussed in
the introduction.  Let us verify that the tension of the domain wall,
compared to $(M_{4})^3$, can be lowered by tuning the volume modulus
$V \sim R^6$. In the estimate below, we set $g_s \sim O(1)$.  A
5-brane wrapping a three cycle of size $R^3$ gives rise to a domain
wall tension
\eqn\valsig{\sigma \sim R^3/(\alpha^{\prime})^3.}
On the other hand, $M_{4} \sim ({\alpha^{\prime}})^{-2} R^3$, so that
$\sigma/(M_{4})^3 \sim (\alpha^{\prime})^3/R^6$.  This ratio can be
made small by taking $R$ to be large.\foot{The tension of the domain
wall depends on both the volume and the moduli that control the sizes
of the relevant three cycles.  We will show that the backreaction of
the wall on all moduli, including these, is small.}

The rest of this section is organised as follows.  The time dependent
dynamics of the domain wall is analysed in \S3.2, under the assumption
that the the wall moves as a single cohesive unit, driven primarily by
its net tension.  A specific example of two vacua and the
interpolating domain wall is discussed next, in \S3.3.  Finally in
\S3.4, the relative forces between branes which make up the wall are
analysed.  These forces are found to be small, thereby justifying the
analysis of \S3.2.

\subsec{Bubble dynamics}

We begin by neglecting the backreaction of the domain wall on the
metric and other closed string modes, and analyse its trajectory in
flat space. Next, we estimate the backreaction effects and show that
they are small most of the time.  All along we work with walls of
tension $\sigma \ll (M_{4})^3$.

A spherical  domain wall in flat space is described by the action,
\eqn\one{S=-\int_{t_i}^{t_f} dt
~4\pi \sigma \rho^2 \sqrt{1-{\dot {\rho}}^2},}
or equivalently an energy
\eqn\two{M={4\pi \sigma \rho^2 \over \sqrt{1-{\dot\rho}^2}}.}

The dynamics is easy to work out in detail.  For fixed $M$ and initial
outward radial velocity, the bubble expands to a maximum size
\eqn\rhomax{4 \pi \sigma \rho_{max}^2=M,}
then recollapses.\foot{We do not consider trajectories where the wall
moves in the internal directions. This is a consistent approximation
to make.}

Birkhoff's theorem tells us that the spherically symmetric geometry
outside the wall is described by the Schwarzschild metric, while that
inside the wall is described by flat space.  The Schwarzschild radius
$R_s \sim G_N M$ (with $G_N \sim M_{4}^{-2}$ the 4d Newton's
constant).  The gravitational backreaction is therefore small as long
as
\eqn\condga{\rho \gg G_N M.}
When $\sigma/(M_{4})^3 \ll 1$, \condga\ can be met by suitably
choosing the initial radius $\rho_i$ and the total energy of the wall.
E.g., for a slowly moving wall, ${\dot {\rho}} \ll 1$, \condga\ is met
by taking,
\eqn\condgb{\rho_i \ll {(M_{4})^2 \over \sigma} 
\sim {R^3 \over \alpha^{\prime}}.}
where we have used \two\ and \valsig.  Ultimately, as the bubble
recollapses, \condga\ will no longer hold and the gravitational
backreaction will get significant, potentially leading to the
formation of a black hole.\foot{See however the discussion of
stability in \S3.4}

The important thing to emphasise is that even if a black hole
eventually forms, by tuning the volume and other moduli, the time for
which the wall lies outside the black hole horizon can be made as
large as one wishes.  E.g., the time it takes starting from an initial
radius $\rho_i \le {R^3 \over \alpha^{\prime}}$ to recollapse back to
$\rho_f=\rho_i$ is of order $\Delta t \sim {R^3 \over
\alpha^{\prime}}$.

The domain wall also acts as a source for the various moduli that
determine its tension. We now show that the back reaction on these
moduli is also small as long as the domain wall is well outside its
Schwarzschild radius.  We denote the canonically normalised modulus
under consideration as $\psi$, and by an additive shift ensure that
asymptotically far away, $r \rightarrow \infty$, $\psi =0$ (e.g.\ for
the radius $\psi \sim (\log(R) -\log(R_{\infty}) )$.  One can show
that $\psi$ satsifies the equation:
\eqn\bcoa{\nabla^2 \psi 
= {\beta \sigma \over (M_{4})^2 } \ \  \sqrt{1-{\dot \rho(t)}^2} \ \
 \delta(r-\rho(t)).}
The right hand term arises because the tension, $\sigma$, depends on
$\psi$.  $\beta$ is determined by this dependence, and $\rho(t)$ is
the radius of the wall.

Since our main concern is the part of the trajectory where the bubble
is well outside its Schwarzschild radius, we consider a simplified
model for the domain wall's history below.  We assume the wall is
constructed at time $t=t_i$ and then evolves till $t=t_f$ with the
radius, $\rho(t)$, meeting the condition \condga\ all along. At time
$t_f$ we assume the bubble is destroyed.

In this example, $\psi$ satisfies the following boundary
conditions:\foot{ $r,t$ are the usual radial and time coordinates in
flat 4d space.}  it vanishes as $r \rightarrow \infty$ for all $t$,
and also as $t \rightarrow \pm \infty$ for all $r$.  Also, we choose
boundary conditions such that $\psi = 0$ inside the bubble.

The resulting solution for $\psi$ is,
\eqn\solaps{\psi={f_{+}(t+r) + f_{-}(t-r) \over r}. }
$f_{\pm}$ meet two junction conditions across the wall:
$f_+(t+\rho(t)) + f_-(t-\rho(t)) = 0,$ and, ${f_+'(t+\rho(t))+
f_-'(t-\rho(t))} = {\beta\,\sigma_0 \rho(t) \over M_4^2},$ with prime
indicating derivative with respect to argument.  $\sigma_0$ is the
tension at $\psi=0$.  Using these, we can solve for $f_{-}$ in terms
of the trajectory $\rho(t)$ :
\eqn\finalvalfm{f_-(t-\rho(t))=
\left\{\matrix{\strut 0 & t < t_i \cr
\strut {\beta\,\sigma_0 \over 2 M_4^2} \int_{t_i}^{t_{\phantom f}}
dt \rho(t) \bigl(1-\dot\rho(t)^2\bigr) & t_i < t < t_f \cr
\strut {\beta\,\sigma_0 \over 2 M_4^2}\int_{t_i}^{t_f} dt \rho(t)
\bigl(1-\dot\rho(t)^2\bigr) & t_f < t}
\right. .}
$f_+$, and finally $\psi$ can then be determined from \solaps\ and the
junction conditions above.  A small backreaction means $\psi \ll
1$. It is easy to see from \finalvalfm, \solaps\ that this requirement
is met when the bubble radius is much larger than the Schwarzschild
radius, \condga.

\subsec{An Example}

As a concrete example, consider a spherical bubble of the standard
${\cal N}=4$ vacuum inside the ${\cal N}=2$ vacuum determined by
\fluxcho.

The domain wall in this case consists of two kinds of D5 branes and
two kinds of NS5 branes.  The D5 branes wrap the three cycles $x_1 =
x_2 = y_3 = 0$ and $y_1 = y_2 = y_3 = 0$, respectively, with
appropriate orientations.  Each of these branes carries two units of
D5 brane charge.  The NS5 branes, each carrying two units of NS
5-charge, wrap the three cycles $x_1=x_2=x_3=0$ and $y_1 = y_2 = x_3 =
0$ respectively.

The compactification also has $64$ O3 planes.  Finally, the ${\cal
N}=4$ vacuum has $16$ D3 branes, while the ${\cal N}=2$ vacuum has
$12$ D3 branes \tadpole.  The extra D3 branes in the ${\cal N}=4$
vacuum terminate on the $5$ branes.

\subsec{Stability}

Our discussion of the wall dynamics assumed that the different branes
making up the wall do not come apart due to relative forces between
them.  This assumption is worth examining, since the configuration
breaks supersymmetry and $g_s \sim O(1)$ in these vacua.

To begin, it is useful to understand the two sources of susy breaking
in this configuration.  First, there is the curvature of the two
sphere in spacetime, which via the bubble tension gives rise to
collective motion of the branes.  Second, there is the presence of
both the branes, and the three-form flux.

To understand the second source, it is helpful to study the example of
\S3.3.  Here, we take the decompactification limit, $R \rightarrow
\infty$, and consider planar parallel branes in $R^{3,1}$ in this
limit (while keeping the orientation of the branes in the internal
directions unchanged).  The spinor conditions can be analysed as in
\refs{\hanany,\GPOne}. One finds that the configuration of branes and
O3 planes of \S3.3\ preserves ${\cal N}=1$ susy, i.e. four
supercharges.  Also, it turns out that any two components, e.g., two
kinds of branes or one brane and the O3 planes, preserve ${\cal N}=2$
susy. Breaking to ${\cal N}=1$ requires three kinds of
branes/planes.\foot{ Essentially the same analysis of susy breaking
and stability applies to the domain wall obtained by replacing the
${\cal N}=2$ vacuum \fluxcho\ with the example in \S7.1\ of \kst. This
latter ${\cal N}=2$ vacuum lifts all complex structure moduli.}

As $R \rightarrow \infty$, the effect of the flux $G$ vanishes and can
be neglected.  But for finite $R$, the $G$ flux \fluxcho\ contributes
additional terms in the spinor equations \refs{\GPOne,\BB}.  Now it
turns out that the spinor conditions imposed by the brane
configuration are in conflict with those imposed by the fluxes.  As a
result, supersymmetry is completely broken at finite $R$.

With this example in mind, let us return to the general discussion
about relative forces between branes.  These are of two kinds.  First,
the different branes couple to different ambient fluxes (the effect of
the flux sourced by a brane should be neglected in this interaction).
The electric potential energy of a brane in the flux background is
\eqn\VfluxI{V_{\rm flux}(\rho) \sim \mu \int
C_{(6)} \sim (\alpha^{\prime})^{-2} \rho^3, }
where $C_{6}$ is the appropriate gauge potential and we have set $g_s
\sim O(1)$.  $\mu$ is order one in string units.  The energy in the
tension, \valsig, is $V_{\rm tension} \sim \sigma \rho^2 \sim R^3
\rho^2/(\alpha^{\prime})^3$.  Comparing, we see that
\eqn\Vratio{V_{\rm flux}(\rho)/V_{\rm tension}(\rho) \sim
\alpha^\prime \rho /R^3.}
This ratio is small as long as the bubble radius is bigger than the
Schwarzschild radius, \condga.

Second, interbrane forces could arise if in the absence of flux, the
brane/plane configuration breaks susy completely, or, as in the
example above, the brane/plane configuration preserves only ${\cal
N}=1$ susy, which allows for a superpotential to be generated.  One
expects the resulting (super)potential to scale like the common world
volume of the branes required to reduce the susy to ${\cal N}\le 1$.
To be comparable with  $V_{\rm tension}$ the potential must scale
like $R^3$, so all the required branes must be parallel in the
internal direction.\foot{In \S3.3\ three branes/planes are neccessary
to break susy to ${\cal N}=1$, and the common world volume lies
entirely in the $R^{3,1}$ spacetime.  Thus the resulting potential energy $\delta V$
scales like $\delta V \sim \rho^2/(\alpha^{'})^3$ and  is small compared
to  $V_{\rm tension}$.
}  ${\cal N} \le 1$ susy then
leaves only one possibility: a pair of NS and D5 branes. Such a pair
of 5 branes breaks all susy's.  However, in this case the pair can be
replaced by a 5 brane susy preserving bound state carrying both NS and D5 brane
charge, which results in the same jump in $F,H$ flux.  Thus, by
appropriately choosing the components of the domain wall, such forces
can be made small.

\newsec{Discussion}

The general idea of unifying vacua through vacuum bubbles of some
fixed tension, $\sigma$, was discussed by Banks in \banks.  There, the
focus was on how gravitational back-reaction makes it difficult to
imagine using such bubbles as a diagnostic in gravity theories (as
opposed to field theories).  In our construction, however, the moduli
spaces we are unifying allow us to make $\sigma$ very small in 4d
Planck units, and hence to manufacture bubbles which are large but are
not yet black holes.  In such a circumstance, we find the notion of
``unification through vacuum bubbles'' meaningful, even in the
presence of gravity.

Our construction connects vacua with large enough volume.  Starting
with a vacuum where the volume modulus is small, one can imagine first
creating a large region of spacetime where the volume modulus is
large.  This can be done by a slowly varying, large amplitude wave of
the volume modulus. In this region the bubble construction could then
proceed as before. While we have not explored such time dependent
solutions with moduli waves and spherical bubbles in detail, it seems
quite reasonable that they exist.\foot{ A similar construction should
also apply to directly connect two vacua in which some of the moduli
are fixed to different values. The fluxes give rise to a potential on
moduli space, $V_{pot}/(M_{4})^4 \sim (\alpha^{\prime})^6/R^{12}$,
which is small for large $R$.  Hence, the resulting bubble should
still be accessible from the outside vacuum.}

Moving beyond, as a next step in making our construction useful in the
overall scheme of string duality, it would be important to find
transitions connecting e.g.\ ${\cal N}=2$ models on $T^6/Z_2$ to
${\cal N}=2$ compactifications on some more generic Calabi-Yau space.
In such a case, up to standard dualities, one would have successfully
unified the heterotic string on $T^6$ with the best-understood web of
vacua with less supersymmetry.

It would also be interesting to study connectedness by asking if one
can find time dependent solutions which roll between vacua separated
by a finite potential barrier. The fact that $\sigma \ll M_{4}^3$ in
our construction is suggestive.  However, the existence of such
solutions cannot be explored in low energy field theory, since the
fields which create the five-branes would also have to be excited.
Exploring such solutions in string field theory seems difficult, at
the moment.


\centerline{\bf{Acknowledgements}}

We would like to thank A.~Dabholkar, S.~Das, M.~Douglas, S.~Ferrara,
S.~Giddings, M.~Porrati, S.~Shenker, E.~Silverstein, L.~Susskind,
P.~Tripathi, H.~Verlinde, E.~Witten and especially T.~Banks for
helpful and encouraging discussions.  This work was supported in part
by the Department of Energy under contract DE-AC03-76SF00515.  The
work of S.K. is also supported by a Packard Fellowship, a Sloan
Fellowship, and NSF grant PHY-0097915.

\listrefs
\bye